# Solution Processed Large-scale Multiferroic Complex Oxide Epitaxy with Magnetically Switched Polarization

Cong Liu[1,†], Feng An[1,2,†], Paria S.M. Gharavi[3], Qinwen Lu[4], Chao Chen[5], Liming Wang[6], Xiaozhi Zhan[6], Zedong Xu[7], Yuan Zhang[2], Ke Qu[1,8], Junxiang Yao[1], Yun Ou[1,9], Xiangli Zhong[2], Dongwen Zhang[10], Nagarajan Valanoor[3], Lang Chen[7], Tao Zhu[6,11], Deyang Chen[5], Xiaofang Zhai[4], Peng Gao[8], Tingting Jia[1*], Shuhong Xie[2*], Gaokuo Zhong[1*], Jiangyu Li[1,12*]

[1] Shenzhen Key Laboratory of Nanobiomechanics, Shenzhen Institutes of Advanced Technology, Chinese Academy of Sciences, Shenzhen, China.
[2] School of Materials Science and Engineering, Xiangtan University, Xiangtan, China.
[3] School of Materials Science and Engineering, UNSW Sydney, Sydney, Australia.
[4] Hefei National Laboratory for Physical Sciences at Microscale and Department of Chemical Physics, University of Science and Technology of China, Hefei, China.
[5] Institute for Advanced Materials and Guangdong Provincial Key Laboratory of Optical Information Materials and Technology, South China Academy of Advanced Optoelectronics, South China Normal University, Guangzhou, China.
[6] Dongguan Neutron Science Center, Dongguan 523803, China.
[7] Department of Physics, Southern University of Science and Technology, Shenzhen, China.
[8] International Center for Quantum Materials, and Electron Microscopy Laboratory, School of Physics, Peking University, Beijing, China.
[9] Hunan Provincial Key Laboratory of Health Maintenance for Mechanical Equipment, Hunan University of Science and Technology, Xiangtan, China.
[10] Department of Physics, College of Science, National University of Defense Technology, Changsha, China.
[11] Beijing National Laboratory for Condensed Matter Physics and Institute of Physics, Chinese Academy of Sciences, Beijing, China, and Songshan Lake Materials Laboratory, Dongguan Neutron Science Center, Dongguan, China.
[12] Department of Mechanical Engineering, University of Washington, Seattle, United States.

[†] These authors contributed equally to this work.
[*] To those correspondence should be addressed to: jjli@uw.edu; gk.zhong@siat.ac.cn; shxie@xtu.edu.cn; tt.jia@siat.ac.cn

## Abstract

Complex oxides with tunable structures have many fascinating properties, though high-quality complex oxide epitaxy with precisely controlled composition is still out of reach. Here we have successfully developed solution-based single crystalline epitaxy for multiferroic $(1-x)BiTi_{(1-y)/2}Fe_yMg_{(1-y)/2}O_3$-$(x)CaTiO_3$ (BTFM-CTO) solid solution in large area, confirming its ferroelectricity at atomic-scale with a spontaneous polarization of 79~89μC/cm². Careful compositional tuning leads to a bulk magnetization of 0.07±0.035$\mu_B$/Fe at room temperature, enabling magnetically induced polarization switching exhibiting a large magnetoelectric coefficient of 2.7-3.0×10⁻⁷s/m. This work demonstrates the great potential of solution processing in large-scale complex oxide epitaxy and establishes novel room-temperature magnetoelectric coupling in epitaxial BTFM-CTO film, making it possible to explore a much wider space of composition, phase, and structure that can be easily scaled up for industrial applications.

Complex oxides with tunable compositions and structures have fascinating properties including high-temperature superconductivity[1], colossal magnetoresistance[2], superior piezoelectric effect[3], and room-temperature magnetoelectric coupling[4,5], and high-quality single crystalline epitaxial films are essential for exploring their fundamental sciences and technological applications[6]. The composition of complex oxides, however, makes such epitaxial growth challenging via conventional physical vapor depositions (PVD) [7,8], and there is strong desire to develop alternative strategies enabling complex oxide epitaxy. This is particularly important for room-temperature multiferroics that often requires sophisticated compositional engineering[9-12], for example to twist the antiferromagnetic ordering of bismuth ferrite (BFO) into ferromagnetic one, as recently demonstrated in the solid solution of (1-x)BiTi$_{(1-y)/2}$Fe$_y$Mg$_{(1-y)/2}$O$_3$-(x)CaTiO$_3$ (BTFM-CTO) ceramics*(11)*. High-quality epitaxy for oxides with composition as complex as BTFM-CTO, however, remains out of reach. Here we develop sol-gel based solution processing*(14)* for complex oxide epitaxy, with which we have successfully achieved large-scale single crystalline epitaxial BTFM-CTO films with room-temperature multiferroicity and magnetically switched polarization that can be easily scaled up for industrial applications.

We choose BTFM-CTO as our model system for its complex composition, whose baseline structure can be viewed as that of BFO*(4)* (Fig. 1A). The cycloidal spin structure of BFO*(15)* can be disrupted by the solid solution of BTFM-CTO*(11,16)*, resulting in magnetic percolation and bulk magnetization at y>0.6 in BTFM-CTO ceramics*(11)* (Fig. S1A in the Supplementary Materials, SM). In order to enable epitaxial growth of BTFM-CTO films, sol-gel based two-step solution processing (Fig. S1B) has been developed using SrTiO$_3$/La$_{0.7}$Sr$_{0.3}$MnO$_3$ (STO/LSMO) (Fig. 1B) and Nb-doped SrTiO$_3$ (NSTO) substrates, wherein the misfit strain is calculated to be -1.7%*(11)*. The processing is optimized at x=0.14 near morphotropic phase boundary (MPB) and y=0.8 for magnetic percolation, and concentration of the solution is kept low to mitigate the evaporation rate during gelation, with propionic anhydride added to dehydrate the water in the solution and tunes its viscosity*(8)*. As a result, high-quality epitaxial films with typical size up to 20×20 mm$^2$ (Fig. S1C) and atomic smooth surface have been obtained (Fig. S1D), exhibiting root mean square roughness as small as 115 pm.

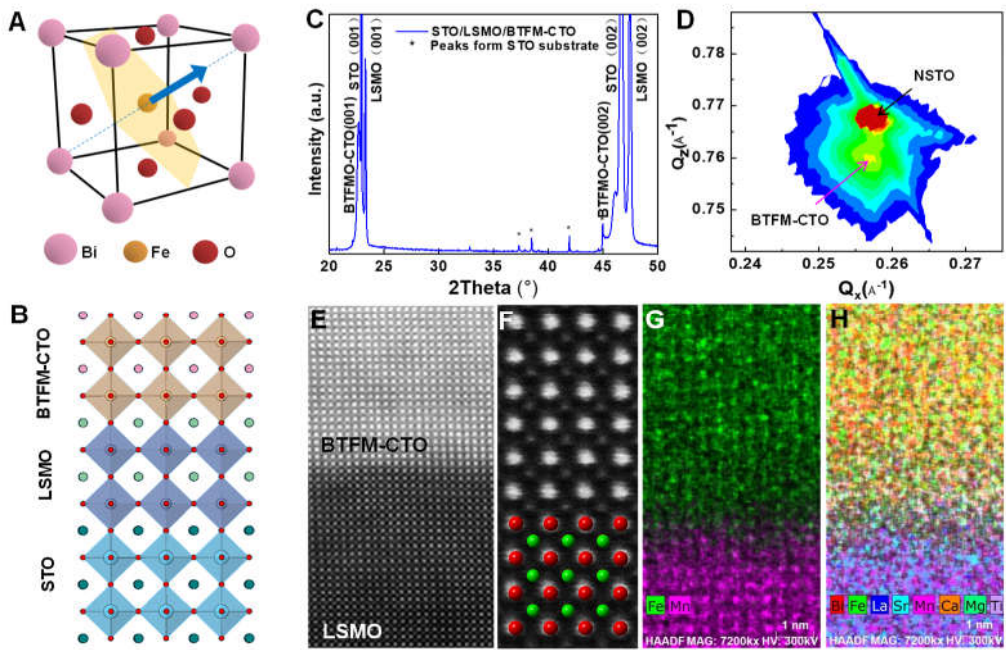

**Fig. 1 Epitaxial structure of BTFM-CTO.** (A) (001)$_{pc}$-oriented BiFeO$_3$ crystal structure with ferroelectric polarization (blue arrow) and antiferromagnetic plane (shaded plane). (B) Heterostructure of epitaxial STO/LSMO/BTFM-CTO. (C) XRD pattern of STO/LSMO/BTFM-CTO. (D) RSM of NSTO/BTFM-CTO at (103) peak of NSTO. (E) Atomic resolution HAADF image at LSMO/BTFM-CTO interface viewed along [010]. (F) Magnified HAADF image of BTFM-CTO. (G-H) Atomic-scale EDS mapping of LSMO/BTFM-CTO interface.

X-ray diffraction (XRD) ω-2theta pattern of STO/LSMO/BTFM-CTO (Fig. 1C) and NSTO/BTFM-CTO (Fig. S2) both demonstrate that films are epitaxially grown along pseudocubic (001)$_{pc}$ direction with no detectable secondary phase. The reciprocal space map (RSM) of NSTO/BTFM-CTO measured around (103) diffraction peak of STO (Fig. 1D) shows that the film possesses identical in-plane lattice parameter as the substrate, measured to be a=3.905 Å and c= 3.945 Å in comparison to 3.962 Å and 3.935 Å reported for BTFM-CTO ceramics*(11)*. Low-magnification cross-sectional scanning transmission electron microscopy (STEM) of STO/LSMO/BTFM-CTO (Fig. S3A-C) suggests a film thickness of 30.14 nm, while quantitative elemental analysis based on energy dispersive x-ray spectroscopy (EDS) mappings (Fig. S3D) reveals uniform distribution of all the elements. The atomically resolved high-angle annular dark field (HAADF) image (Fig. 1E) and selected area electron diffraction (SAED) pattern (Fig. S3E) confirms the high-quality epitaxy at atomic scale near LSMO/BTFMO-CTO interface, and the magnified HAADF image of BTFM-CTO (Fig. 1F) matches well with the lattice structure shown. High-resolution EDS maps (Fig. 1G-H) suggest that the interdiffusion of cations at LSMO/BTFM-CTO interface is minor, limited to only a couple of unit cells. This set of data thus firmly establish high-quality epitaxial growth of BTFM-CTO single crystalline films on both STO/LSMO and NSTO substrates.

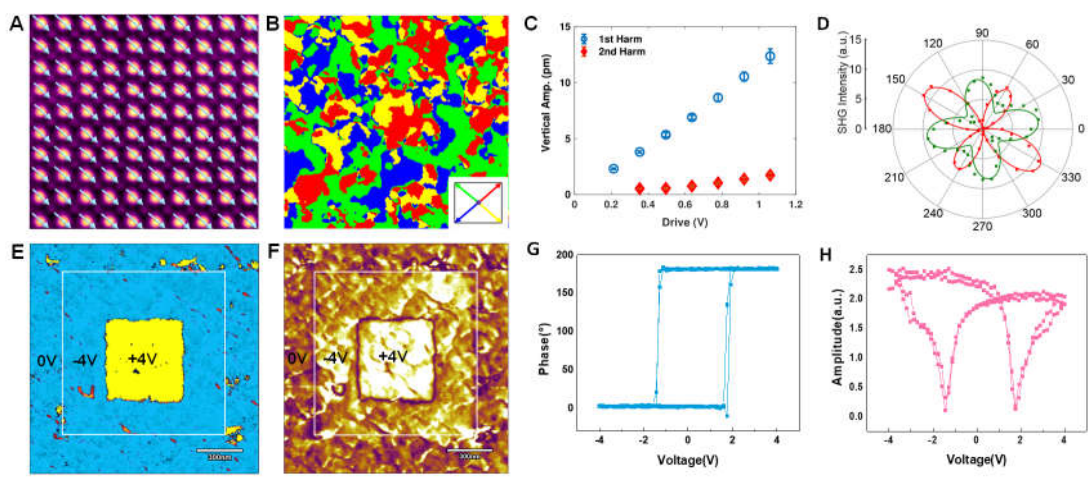

**Fig. 2 Ferroelectricity of STO/LSMO/BTFM-CTO.** (A) Atomic-scale polarization vector of BTFM-CTO overlaid on a cross-sectional HAADF-STEM image. (B)

Mapping of in-plane polarization. (C) Comparison of first and second harmonic piezoresponses versus driving voltage. (D) Measured SHG signal represented as a polar diagram, where the green and red squares are the *p*- and *s*-polarized SHG signals, and the green and red lines are the corresponding fittings. (E-F) PFM phase and amplitude mappings of electrically poled domains. (G-H) PFM phase hysteresis and amplitude butterfly loops.

At atomic scale, the polar order of BTFM-CTO is determined from the displacement of the B-site cation to the center of the surrounding A-site cations *(17)* using the atomically resolved HAADF image (Fig 2A, Fig. S4), and the overlaid polarization vector is found to be uniformly distributed along $[111]_{pc}$ direction. The magnitude of polarization is calculated to be 79~89 $\mu C/cm^2$, comparable to 98 $\mu C/cm^2$ reported for $(001)_{pc}$-oriented BFO *(18)* and larger than ~50 $\mu C/cm^2$ measured for BTFM-CTO ceramics and polycrystalline film *(11,19)*. At mesoscale, lateral piezoresponse force microscopy (LPFM) mappings before and after 90° sample rotation are carried out (Fig. S5) and then combined into in-plane polarization mapping (Fig. 2B), revealing irregular domain pattern with domain closure configuration. The corresponding vertical PFM (VPFM) mappings exhibit no phase contrast (Fig. S6), and thus domain walls are determined to be 71° type *(20)*. Point-wise first and second harmonic vertical piezoresponses acquired under a series of excitation voltages (Fig. 2C) reveal that the electromechanical response is predominantly linear piezoelectric *(21)*, consistent with the presence of strong polarization, which is also confirmed at the macroscopic scale by the polarization-dependent *p*- and *s*-polarized second harmonic generation (SHG) signals*(22)* (Fig. 2D). The polarization can be switched by an external electric field, as demonstrated by VPFM phase and amplitude mappings after box-in-box poling*(23)* by ±4 V (Fig. 2E-F, Fig. S7) as well as classical hysteresis and butterfly loops (Fig. 2G-H) acquired from point-wise switching spectroscopy*(24)*. This set of data thus establish ferroelectricity of epitaxial BTFM-CTO film.

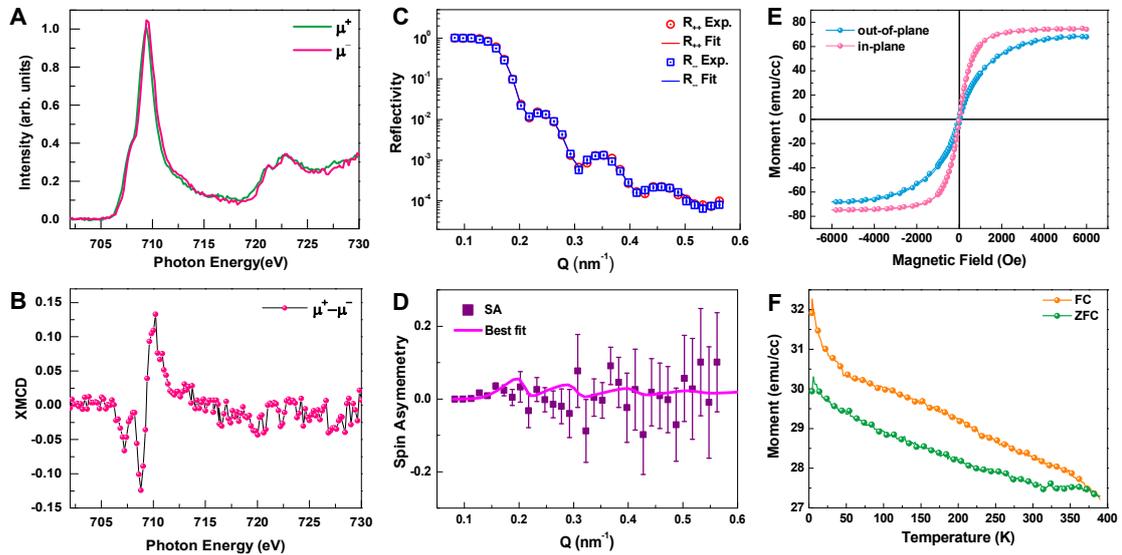

**Fig. 3 Magnetism of NSTO/BTFM-CTO.** (A) XAS spectra at the Fe $L_{2,3}$ edge under 4500 Oe. (B) XMCD calculated from A. (C) PNR with the spin-dependent neutron reflectivities $R_{++}$ and $R_{--}$ at room temperature under a 7000 Oe magnetic field applied along the in-plane direction. (D) Spin asymmetry (SA) calculated from C. (E) out-of-plane (blue) and in-plane (pink) magnetization curve versus applied magnetic field at room temperature. (F) ZFC (green) and FC (orange) temperature dependence of magnetization.

X-ray absorption spectrum (XAS) of NSTO/BTFM-CTO is measured under a magnetic field of 4500 Oe with left- and right-hand circular lights, both of which exhibit the features corresponding to $Fe^{3+}$*(25,26)*, with a left shoulder at ~708 eV and a peak at ~709.5 eV on the $L_3$ edge (Fig. 3A). The X-ray magnetic circular dichroism (XMCD) signal is obtained by calculating the difference between the left- and right-hand XAS signals, showing strong dichroism on the $L_3$ edge and non-observable dichroism on the $L_2$ edge (Fig. 3B). From the spectra, a spin moment of ~0.08 $\mu_B$/Fe and an orbital moment of ~0.05 $\mu_B$/Fe are calculated, similar to 0.03 $\mu_B$/f.u. reported for Co-doped BFO*(27)*. We also use polarized neutron reflectometry (PNR) to examine NSTO/BTFM-CTO, which is a powerful tool for investigating the depth-resolved magnetization profile of thin films that is exclusively sensitive to the long-range order and thus can eliminate the signals resulted from magnetic contamination or cluster*(28)*. The non-spin-flip specular reflectivities of polarized neutrons ($R_{++}$ and $R_{--}$, Fig. 3C) are dependent on the sample magnetic and nuclear scattering length density denoted as mSLD and nSLD (Fig. S8), from which the spin asymmetry $SA(Q) = \frac{R_{++} - R_{--}}{R_{++} + R_{--}}$ as a function of wave vector transfer $Q$ can be calculated (Fig. 3D). Note that mSLD is directly proportional to the magnetization of the sample since it is much smaller than nSLD, and the magnetization trends can be readily extracted by examining the magnitude of the SA features*(28)*. The difference between $R_{++}$ and $R_{--}$ is rather small, and the best fit yields a magnetization of 0.05 ± 0.024 $\mu_B$/f.u., corresponding to 0.07 ± 0.035 $\mu_B$/Fe. This is consistent with the magnetization determined from XMCD and is larger than 0.0097 $\mu_B$/Fe reported for bulk BTFM-CTO ceramics*(9)*. The M-H loop measured at room temperature demonstrates weak ferromagnetism with a saturated magnetic moment of ~72 emu/cm$^3$ at 6000 Oe (Fig. 3E), and remnant magnetization out-of-plane is estimated to be ~3 emu/cm$^3$ (Fig. S9). This saturation value is larger than that estimated from XMCD and PNR, which might be due to the existence of oxygen vacancies and impurities that cannot be detected by PNR*(29,30)*. Splitting of zero-field-cooling (ZFC) and field-cooling (FC) magnetization-temperature (M-T) curves under a detective field of 200 Oe occur at ~370 K (Fig. 3F), consistent with value report in BTFM-CTO ceramics*(11)*. This set of data thus established room-temperature bulk magnetization in epitaxial BTFM-CTO film.

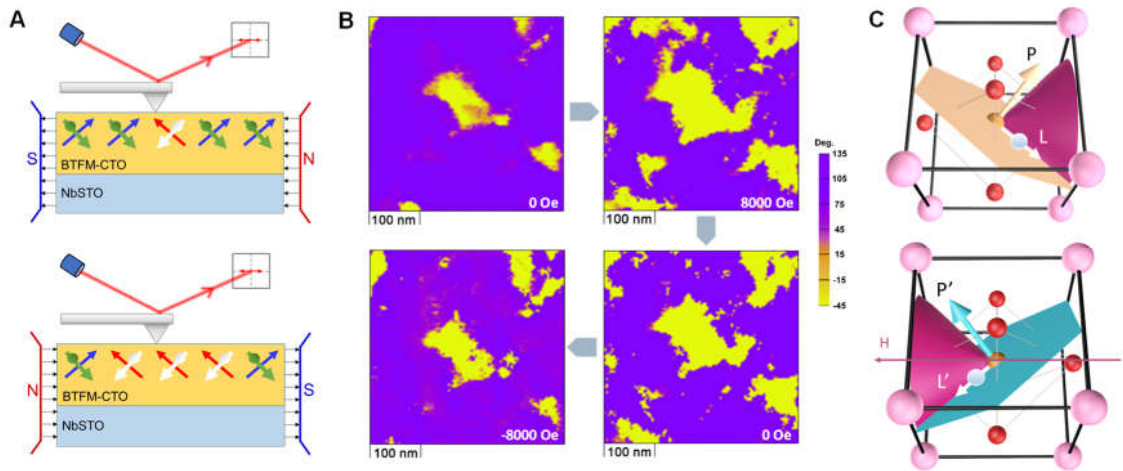

**Fig. 4 Magnetoelectric coupling of BTFM-CTO.** (A) Schematic experimental setup for polarization switching under an in-plane magnetic field; blue and red arrows indicate polarization while green and white arrows indicate net magnetization. (B) LPFM domains expand, maintain and shrink under different in-plane magnetic fields. (C) Polarization switching pathway under the applied magnetic field; P and P' denote

polarization before and after switching, while L and L' represent the corresponding magnetization.

The ferromagnetism in epitaxial BTFM-CTO, in combination with its coupling with polarization, raises an exciting prospect of switching polarization by an external magnetic field at room temperature in a single-phase thin film, and thus we examine LPFM domain patterns of NSTO/BTFM-CTO film under the influence of opposite in-plane magnetic fields *(10,31)*. In-plane domains with 180° phase contrast is evident in the original LPFM mapping in the absence of any magnetic field (Fig. 4A), and upon the application of +8000 Oe field, yellow domains expand at the expense of purple domains (Fig. 4B), while VPFM contrast is intact (Fig. S10). The fraction of the switched polarization is estimated to be 37%, and the magnetoelectric coefficient is calculated to be $\frac{\Delta P}{\Delta H} =$ 2.7-3.0×10$^{-7}$ s/m, in comparison to 1.3×10$^{-7}$ s/m estimated for solid solution of lead zirconium titanate (PZT) and lead iron tantalate (PFT) *(10)*. When this magnetic field is removed, the ferroelectric domains are maintained, suggesting that the magnetically switched polarizations are nonvolatile. A magnetic field of -8000 Oe opposite to the original field then switches the ferroelectric domain back to the original configuration, and similar behavior is also observed in a larger region (Fig. S11A-D) as well as in different samples (Fig. S12). Throughout the processes, the topography is unchanged (Fig. S11E-H). Note that polarization is perpendicular to the easy magnetic plane (Fig. 4C), and thus under an external magnetic field opposite to the magnetization, the magnetic moment rotates around the applied magnetic field, forming a cone with the magnetic field as the axis. This eventually results in flip of magnetization easy plane to reduce the angle between the magnetic field and moment, leading to simultaneously switched in-plane polarization as observed in Fig. 4B.

In summary, we have successfully developed sol-gel based complex oxide epitaxy for multiferroic BTFM-CTO solid solution in large area, enabling convenient compositional tuning in combination with strain engineering for magnetically switched polarization. The process is simple, fast, cost-effective, and can be easily scaled up for industrial applications, making it possible to explore a much wider space of composition, phase, and structure of complex oxides.

**Acknowledgements**

**Funding:** We acknowledge National Key Research and Development Program of China (2016YFA0201001, 2016YFA0401004), National Natural Science Foundation of China (11627801, 51702351, 51672007, 11574287, 11574375, U1832104, 11704130, 51572233, 61601217), Shenzhen Science and Technology Innovation Committee (JCYJ20170818163902553, JCYJ20170413152832151, JCYJ20170307165829951), the Leading Talents Program of Guangdong Province (2016LJ06C372), and the Hong Kong, Macao and Taiwan Science & Technology Cooperation Program of China (2015DFH10200). **Authors Contributions**: The project was conceived by JYL, TTJ, and SHX, and coordinated by JYL. Films were synthesized by CL under the guidance of TTJ, SHX, DYC and assisted by CC, JXY and YO. XRD was carried out and analyzed by CL under the guidance of TTJ and SHX. RSM was carried out and analyzed by QWL, ZDX and CC under the guidance of XFZ and LC. TEM was carried out and analyzed by PSMG, KQ, NV and PG. AFM was carried out and analyzed by FA, JYL and GKZ. SHG was carried out and analyzed by YZ and GKZ with DWZ and XLZ. XMCD was carried out and analyzed by QWL and XFZ. PNR was carried out and analyzed by LMW, XZZ, and TZ. SQUID measurement was carried out and analyzed by ZDX and QWL under the guidance of LC and XFZ.

JYL and CL wrote the manuscript, and all the authors participated discussions and writing. **Competing Interests:** The authors declare no conflict of interests.

**Supplementary Materials**

Materials and Methods

Figs. S1 – S12

References (32 – 33)

# Supplementary Materials for

# Solution Processed Large-scale Multiferroic Complex Oxide Epitaxy with Magnetically Switched Polarization

Cong Liu[1†], Feng An[1,2†], Paria S.M. Gharavi[3], Qinwen Lu[4], Chao Chen[5], Liming Wang[6], Xiaozhi Zhan[6], Zedong Xu[7], Yuan Zhang[2], Ke Qu[1,8], Junxiang Yao[1], Yun Ou[1,9], Xiangli Zhong[2], Dongwen Zhang[10], Nagarajan Valanoor[3], Lang Chen[7], Tao Zhu[6,11], Deyang Chen[5], Xiaofang Zhai[4], Peng Gao[8], Tingting Jia[1*], Shuhong Xie[2*], Gaokuo Zhong[1*], Jiangyu Li [1,12*]

[1] Shenzhen Key Laboratory of Nanobiomechanics, Shenzhen Institutes of Advanced Technology, Chinese Academy of Sciences, Shenzhen, China.
[2] School of Materials Science and Engineering, Xiangtan University, Xiangtan, China.
[3] School of Materials Science and Engineering, UNSW Sydney, Sydney, Australia.
[4] Hefei National Laboratory for Physical Sciences at Microscale and Department of Chemical Physics, University of Science and Technology of China, Hefei, China.
[5] Institute for Advanced Materials and Guangdong Provincial Key Laboratory of Optical Information Materials and Technology, South China Academy of Advanced Optoelectronics, South China Normal University, Guangzhou, China.
[6] Dongguan Neutron Science Center, Dongguan 523803, China.
[7] Department of Physics, Southern University of Science and Technology, Shenzhen, China.
[8] International Center for Quantum Materials, and Electron Microscopy Laboratory, School of Physics, Peking University, Beijing, China.
[9] Hunan Provincial Key Laboratory of Health Maintenance for Mechanical Equipment, Hunan University of Science and Technology, Xiangtan, China.
[10] Department of Physics, College of Science, National University of Defense Technology, Changsha, China.
[11] Beijing National Laboratory for Condensed Matter Physics and Institute of Physics, Chinese Academy of Sciences, Beijing, China, and Songshan Lake Materials Laboratory, Dongguan Neutron Science Center, Dongguan, China.
[12] Department of Mechanical Engineering, University of Washington, Seattle, United States.

† These authors contributed equally to this work.
* To those correspondence should be addressed to: jjli@uw.edu; gk.zhong@siat.ac.cn; shxie@xtu.edu.cn; tt.jia@siat.ac.cn

## 1. Sol-gel precursor preparation and spin coating method

### 1.1 BTFM-CTO Solution preparation

The solution of $0.86BiTi_{3/8}Fe_{2/8}Mg_{3/8}O_3$–$0.14CaTiO_3$ was prepared by two steps. First, sufficiently dissolving the starting materials of bismuth nitrate ($Bi(NO_3)_3 \cdot 5H_2O$, ACS 98%, Aladdin), iron nitrate ($Fe(NO_3)_3 \cdot 9H_2O$, ACS 98.5%, Aladdin), magnesium nitrate (99%, Sigma Aldrich), calcium acetate (99%, Aladdin), and butyl titanate (99%, Aladdin) one by one in 2-methoxyethanol (2-MOE, ACS >99.5%, Aladdin) solvent, with 5% excess bismuth nitrate added to compensate the evaporation of bismuth during the annealing process. Then citric acid (ACS >99.7%, Aladdin) was added into the precursor solution as the chelation agent. Keep the solution in constant stirring for 12 hours to obtain the 0.3 mol/L dense solution. In the second step, 10 mL propionic anhydride (GC >98.5%, Aladdin) was added into the mixture of 20 mL dense solution and 20 mL 2-MOE to dehydrate the water of crystallization from the metal salts, and then stirred for another 12 hours and aged for 72 hours to obtain the diluted solution.

### 1.2 spin-coating and annealing process

The substrates were first cleaned by oxygen plasma for 100 seconds. The solution was then spin-coated onto the substrates sequentially at 600 r/min for 7 seconds and 5000 r/min for 15 seconds. After deposition of each layer, the samples were transferred onto the heating stage with a temperature of 180 °C to accelerate the condensation reaction and then pyrolyzed at 380 °C for 15 minutes to remove the organic frameworks. After three layers of spin-coating and pyrolysis process, the samples were annealed at 800 °C for 15 minutes in a rapid thermal processing furnace under oxygen atmosphere, and a cooling rate of 0.25 °C/s is found to be stable for high-quality crystallization.

**1.3 LSMO buffer layer growth method**

$La_{0.7}Sr_{0.3}MnO_3$ (LSMO) buffer layer on $SrTiO_3$ substrate was deposited by pulsed laser deposition (PLD) at 700 °C in an oxygen ambient of 100 mTorr at a growth rate of ~0.7 Å/sec, with 9600 pulses. The films were cooled at 10 °C/min to the room temperature in an oxygen ambient of 1 atmos.

The high quality epitaxial thin film of 20×20 mm² (Fig. S1C)and very smooth surface with root mean square (RMS) roughness of 115 pm (Fig.S1D) are shown below.

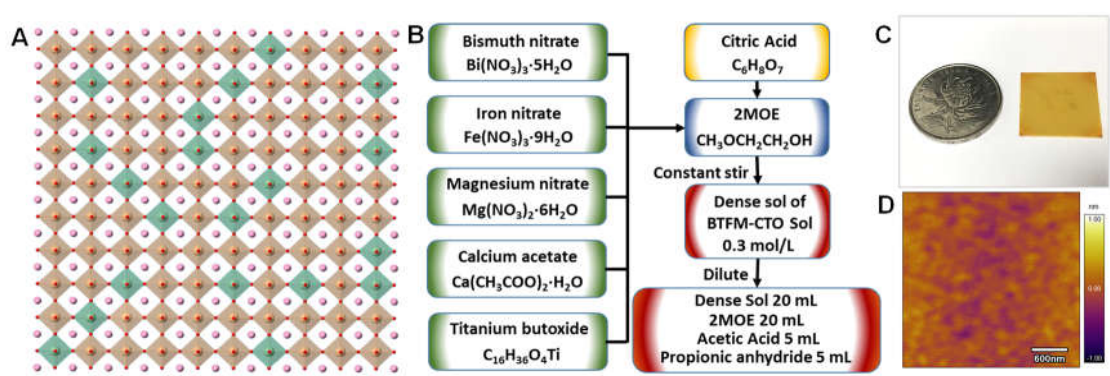

**Fig. S1 Solution processing of BTFM-CTO. (A)** schematic diagram of B-site doping and magnetic percolation for y=0.8. **(B)** Flow chart of precursor solution preparation. **(C)** Optic photo of sol-gel derived 20×20 mm² high quality thin film. **(D)** AFM topography mapping of an ultra-flat NSTO/BTFM-CTO thin film.

## 2. Structure characterization.

### 2.1 X-ray diffraction and reciprocal space mapping

The phase composition was determined by the powder X-ray diffraction (XRD, Bruker D8 Focus X-ray diffraction, Germany) using the Cu K$_\alpha$ radiation (λ=1.5406 Å). The reciprocal space mapping of the thin film was characterized by a 4-circle diffractometer.

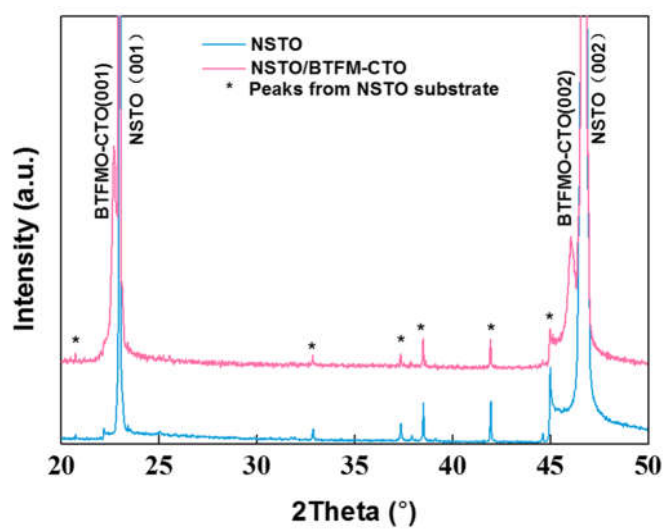

**Fig. S2 XRD pattern of sample NSTO/BTFM-CTO (pink) and NSTO substrate (blue).**

### 2.2 Scanning transmission electron microscopy

Cross-sectional TEM sample of approximately 40 nm thickness was prepared by focused ion beam (FIB) using lift-out technique. High-angle annular dark field (HAADF)-STEM, bright-field (BF)-STEM imaging, and selective area electron diffraction (SAED) patterns were obtained by a Cs-corrected transmission electron microscope (JEOL JEM-ARM200F) operated at 200 kV). High-resolution HAADF STEM images were acquired at an aberration-corrected TEM (FEI Titan Cubed Themis

G2 60-300) operated at 300 kV equipped with an XFEG gun and Bruker Super-X EDX detectors. STEM images were acquired with a beam current of 0.05-0.1 nA, a convergence semi-angle of 25 mrad, and a collection semi-angle snap in the range of 53-260 mrad. The STEM-EDX mapping was obtained with a beam current of 0.2-0.3 nA and counts ranging from 2 k cps to 8 k cps for ~5 minutes.

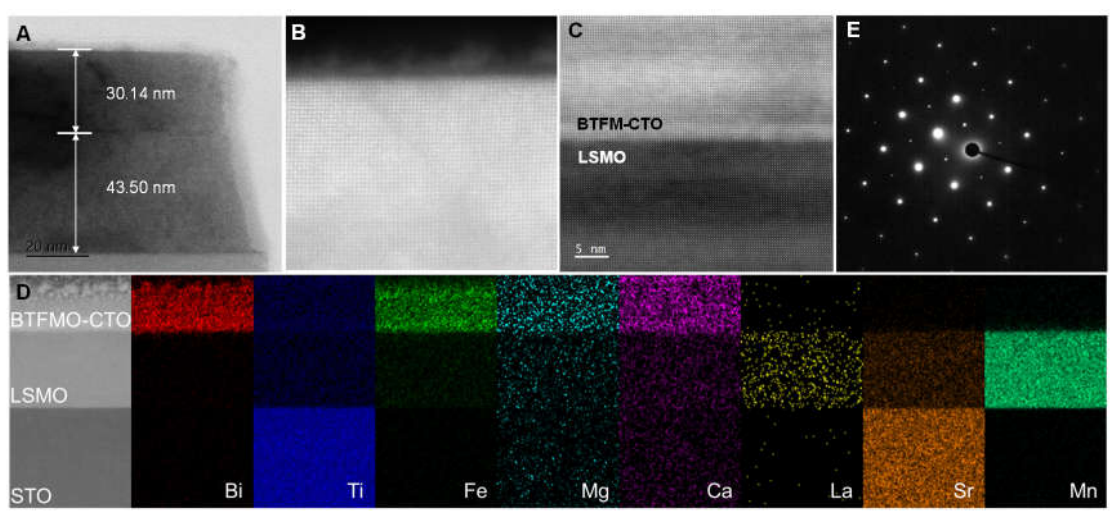

**Fig. S3 STEM, EDS and SAED of STO/LSMO/BTFM-CTO thin film.** **(A)** Low-resolution cross-sectional image. **(B)** Partial-enlarged photo of the surface area. **(C)** Partial enlarged area at the BTFM-CTO/LSMO interface. **(D)** STEM cross-sectional image and corresponding EDS element mapping of the area. **(E)** Selected area electron diffraction pattern of BTFM-CTO/LSMO area.

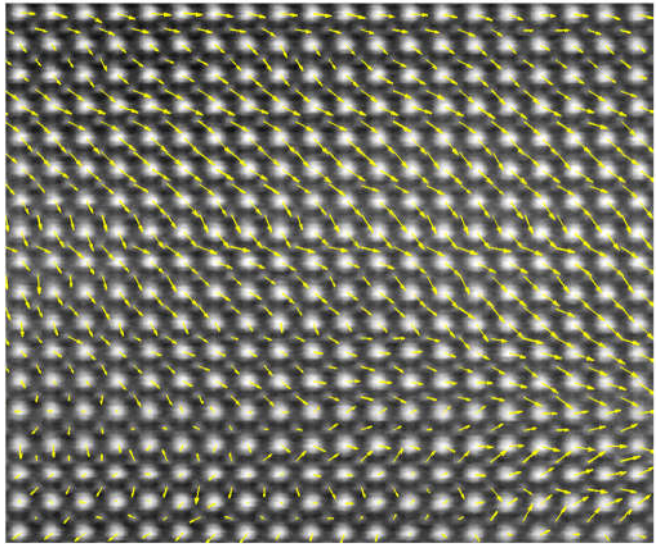

**Fig. S4 Larger region ABSF filtered image of a cross-sectional HAADF-STEM image.**

3. **Ferroelectricity measurement.**

3.1 **Piezoelectric force microscopy measurement**

The lateral and vertical piezoelectric force microscopy (LPFM and VPFM) were performed with an Asylum Research Cypher-ES atomic force microscope (AFM) using dual ac resonance tracking PFM (DART-PFM) by a conductive probe with a spring constant of 2 N/m (FMG-01-Pt) in contact mode with scan rate of 0.5 Hz per line. SSPFM measurement is adopted to illustrate polarization switching by applying a DC voltage up to 4 V, and detected by an ac voltage of 0.5 V at each DC step. Variable field module (VFM) is used to apply an adjustable in-plane magnetic field under AFM.

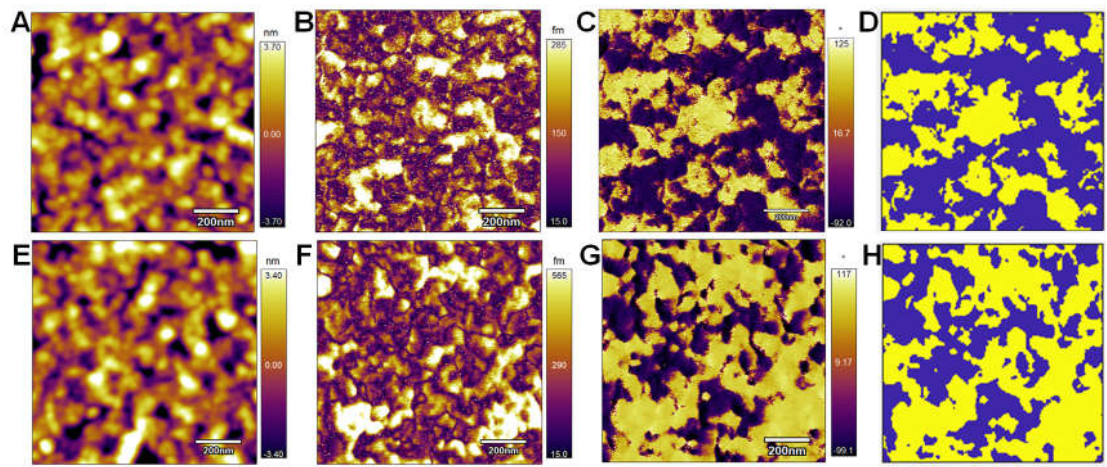

**Fig. S5 LPFM mapping at the same area of STO/LSMO/BTFM-CTO.** Topography, amplitude, phase and binarized phase mapping before (**A-D**) and after (**E-H**) sample rotation of 90°.

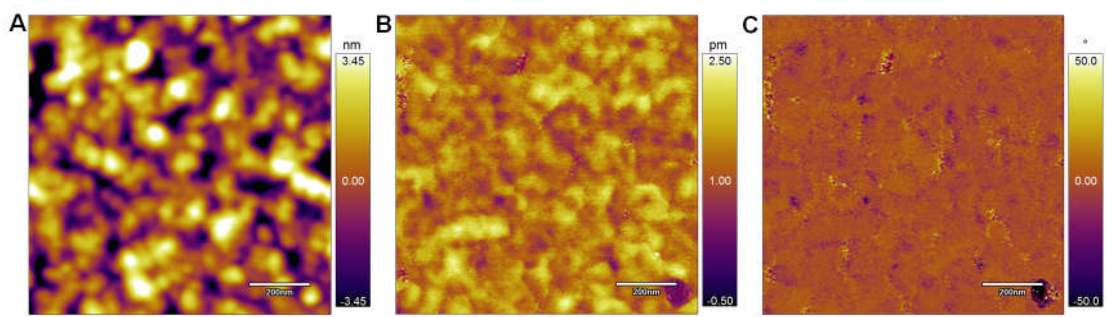

**Fig. S6 VPFM mapping at the same area as Fig. S5. (A)** Topography, **(B)** amplitude and **(C)** phase, demonstrating single-direction vertical polarization.

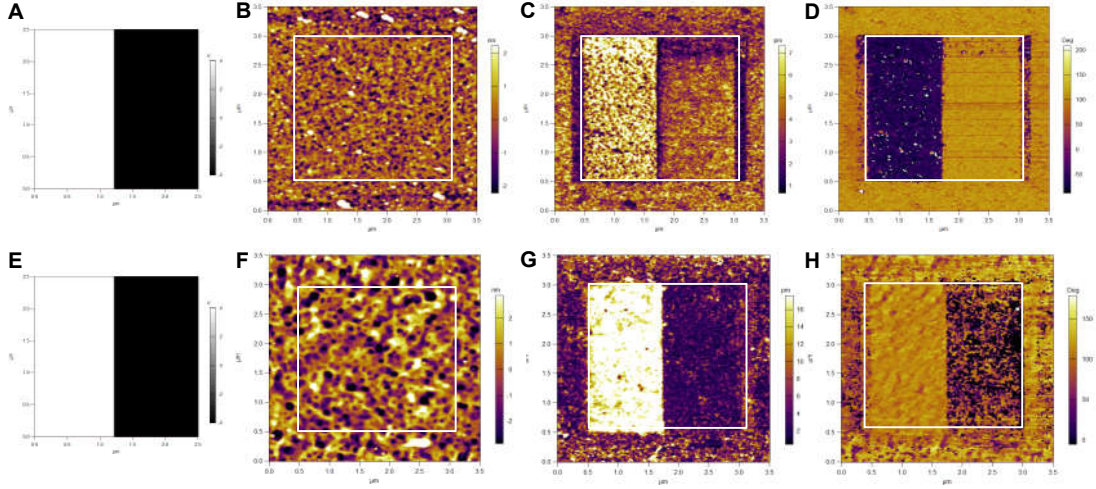

**Fig. S7 VPFM mapping of the thin film after voltage-written.** Bias-voltage pattern and VPFM topography, amplitude and phase mapping of (**A-D**) STO/LSMO/BTFM-CTO in comparison with that of (**E-H**) NSTO/BTFM-CTO. Opposite as-grown vertical polarization directions were due to the self-polling effect by the bottom electrode potential.

## 3.2 Second harmonic generation measurement

The SHG measurements were achieved in reflection configuration with an incident angle of 45°. The 800 nm laser (laser amplifier system, 150 fs, 10 nJ, 76 MHz) was used as incident fundamental beam (electric field $E_\omega$). The sample was installed along STO [100] direction. The polarization-dependent measurements were obtained by rotating the polarization angle of the incident fundamental beam through a half-wave plate, where the 0° or 180° corresponds to *p* polarization ($p_{in}$), and 90° or 270° corresponds to *s* polarization ($s_{in}$), respectively. The *p*-polarized and *s*-polarized SHG signals were analyzed by Glan prisms.

## 4. Magnetism measurement

### 4.1 X-ray magnetic circular dichroism measurement

The XAS and XMCD measurements were performed on beamline BL08U1A at the Shanghai Synchrotron Radiation Facility. The measurements were performed at room temperature and under a vacuum pressure of less than $1\times10^{-6}$ Pa. The XAS data measured using the left- and right-hand circular light were recorded in total electron yield mode (TEY) and normalized by the incident photon flux. The XMCD signal is then calculated as the difference of the left-hand and right-hand XAS signals. The samples were measured with alternating left-polarized (μ+) and right-polarized (μ-) photons in an applied field of 4500 Oe. Both the X-ray beam and the magnetic field were perpendicular to the sample surface.

### 4.2 Polarized neutron reflectivity measurement

Polarized neutron reflectivity (PNR) measurements were carried out at the multipurpose reflectometer (MR) in China Spallation Neutron Source (CSNS) *(32)*. PNR curves were recorded at a magnetic field of 7000 Oe at room temperature, far in excess of the necessary field to saturate the sample. The samples had dimensions of $20\times20$ mm$^2$ on a 0.5 mm thick STO (100) substrate. PNR data were fitted using the default layer model in Genx software package *(33)*. In the model, the volume ratio of BTFM and CTO was set as 0.87 to 0.13, while the literate values of scattering length and density of the composite were calculated from NIST website. The NSTO substrate is treated as slabs with uniform nSLD and zero mSLD. The best fit gives a Factor of

Merit (FOM) of 0.0038.

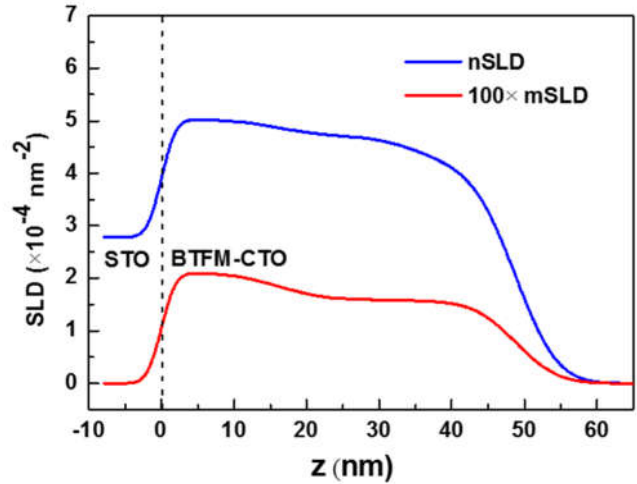

**Fig. S8 Nuclear scattering length density of STO substrate and magnetic scattering length density of BTFM-CTO thin film measured by PNR.**

**4.3 Superconducting quantum interference device measurement**

The M-H (magnetization versus magnetic field) and the M-T (magnetization versus temperature) were measured by a Quantum Design superconducting quantum interference device (SQUID) measurement system (MPMS-7) in the temperature range 10-390 K with the magnetic field (H) applied along the (100) direction of the substrate after cooling down to 10 K in a magnetic field of 6000 Oe and 0 Oe, respectively.

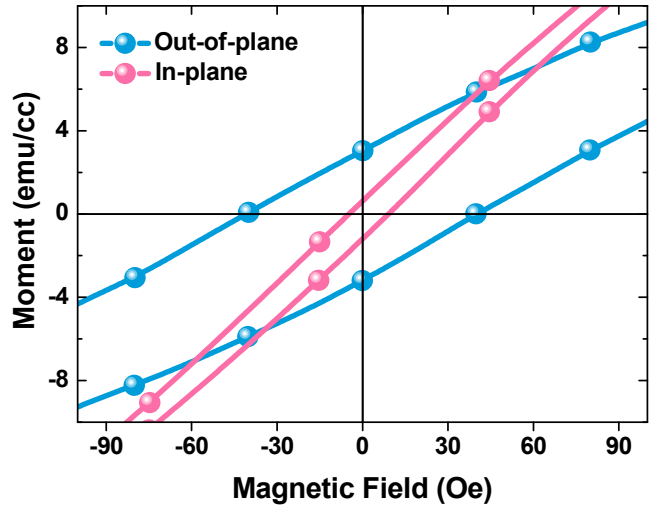

**Fig. S9** Center enlarge of M-H loop in Fig. 4(E).

## 5. LPFM measurement under magnetic filed

The magnetoelectric coupling measurements were operated on a MFP-3D-BIO AFM equipped with the external in-plane variable field module (VFM3, Asylum Research). The sample was placed on VFM3 stage to realize *in-situ* observation of the ferroelectric domain evolution under in-plane different magnetic field.

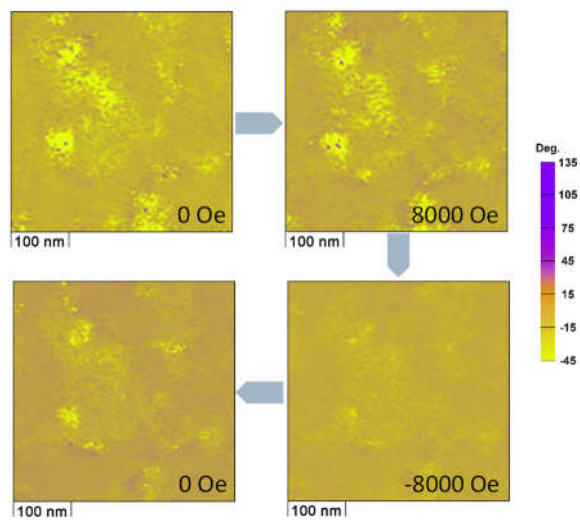

**Fig. S10 VPFM phase mapping under different in-plane magnetic field of the same**

**area as Fig. 5(B).**

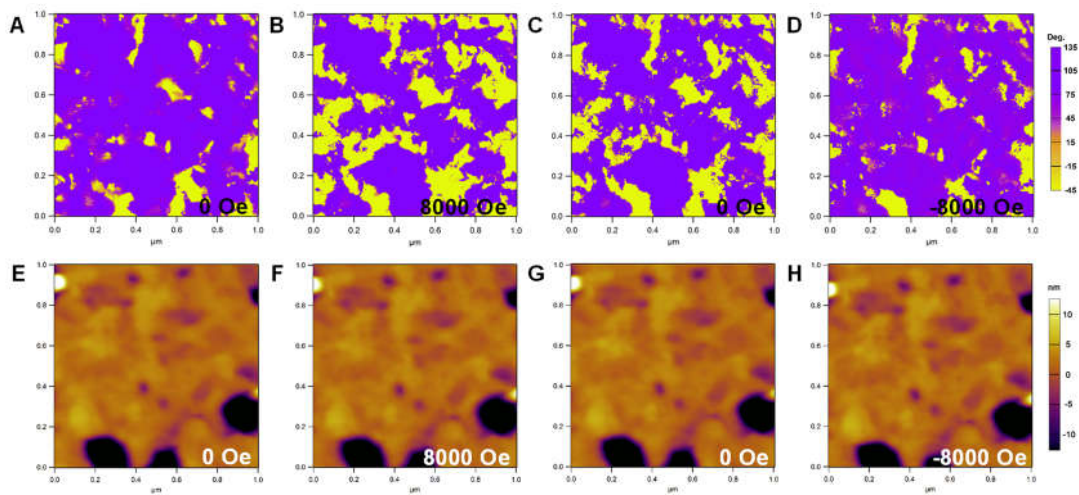

**Fig. S11 Larger region LPFM mapping of NSTO/BTFM-CTO in accordance with Fig. 5b. (A-D)** Phase mapping and **(E-H)** topography of NSTO/BTFM-CTO under different external in-plane magnetic field.

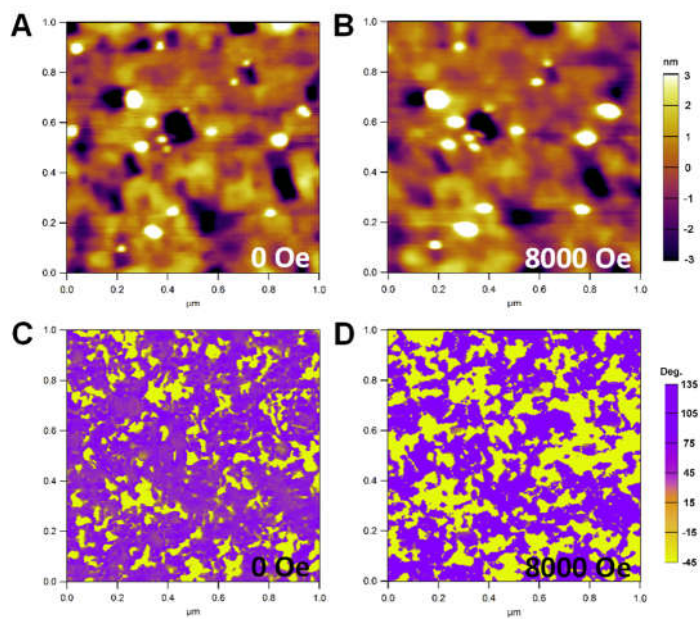

**Fig. S12 LPFM mapping of another NSTO/BTFM-CTO sample. (A-B)** Topography

and (**C-D**) corresponding LPFM phase mapping under different external in-plane magnetic field.